\input harvmac
%
\def\AP#1{Ann.\ Phys. {\bf{#1}}}

\def\NP#1{Nucl.\ Phys. {\bf B{#1}}}
\def\NPs#1{Nucl.\ Phys. {\bf B}(Proc. Suppl.) {\bf #1}}
\def\PL#1{Phys.\ Lett. {\bf B{#1}}}
\def\PLA#1{Phys.\ Lett. {\bf A{#1}}}
\def\PR#1{Phys.\ Rev. {\bf {#1}}}

\def\PRL#1{Phys.\ Rev.\ Lett. {\bf {#1}}}

\input epsf
\def\lsim{\,\,\raise2pt\hbox
   {$\mathop < \limits_{\raise0.5pt\hbox{$\sim$}}$}\,\, }
\def\gsim{\,\,\raise2pt\hbox
   {$\mathop > \limits_{\raise0.5pt\hbox{$\sim$}}$}\,\, }
\Title{KYUSHU-HET-17, SAGA-HE-65}
{\vbox{\centerline{Chiral perturbation theory, finite size effects}
\vskip0.2em\centerline{and}
\vskip0.2em\centerline{the three-dimensional $XY$ model}}}
\centerline{Shin-ichi Tominaga\footnote{*}{ e-mail:tomi1scp@mbox.nc.kyushu-u.ac.jp}
 and Hiroshi Yoneyama$^{\dagger}$\footnote{**}{ e-mail:yoneyama@math.ms.saga-u.ac.jp}}
\bigskip
\centerline{Department of Physics}
\centerline{Kyushu University}
\centerline{Fukuoka, 812 JAPAN}
\bigskip
\centerline{$\dagger$ Department of Physics}
\centerline{Saga University}
\centerline{Saga, 840 JAPAN}
\vskip1cm
\centerline{{\bf Abstracts}}
 We study finite size effects of   the d=3  $XY$ model in terms  of the 
chiral perturbation  theory. We calculate by Monte Carlo simulations physical quantities
  which are, to order of $(1/L)^2$,  uniquely determined only  by  two low energy constants. 
They are the  magnetization and the helicity modulus 
(or the Goldstone boson decay constant) in infinite volume.
We also pay a special attention to  the region of the validity of the two 
possible expansions in the theory.
\par
\Date{}
\newsec{Introduction} 
One has to cope with finite size effects in study of numerical simulations on lattices.
 Recently  Hasenfratz and Leutwyler  \ref\HL{P. Hasenfratz and H. Leutwyler
, \NP{343}, 241 (1990) }
 have  applied to this issue the  chiral perturbation theory \ref\GL{J. Gasser and H. Leutwyler, 
\AP{158},142 (1984)}
 which is   systematic  and  quite alternative to the finite size 
scaling
\ref\FV{M. E. Fisher and V. Privman, \PR{B32}, 447 (1985)}
\ref\BZ{E. Br\'ezin and J. Zinn-Justin, \NP{257}[FS14], 867 (1985)}.
The  chiral perturbation theory is originally a low energy effective theory of QCD and it has
 some systematic expansions in momentum  of  pions. 
For a system in a finite box with  volume  $V=L^d$, which involves  
    Goldstone boson(s), long range properties are dominated by the finite mass of Goldstone
bosons.  By controlling its effect in a way, the chiral perturbation theory enables one 
to calculate
  physical quantities in a systematic series  in terms of $1/L$ .
  As a way of controlling, an external source coupled to the field in consideration
plays an important role.   There are two manners, one is called the $\epsilon$-expansion,
 and another is the  $p$-expansion, which result in the series in different powers of 
$1/L$.  Each of them  has its own  characteristic region  in the space  of  the the external
 source $j$.  The $\epsilon$-expansion  characterizes the domain where $j \sim L^{-d}$,
while the $p$-expansion corresponds to $j \sim L^{-2}$.
\par 
In the present paper we   consider  the $d=3$  dimensional O(2) non-linear sigma model,
 or $XY$-model.
Though this model is simple, it is  a good laboratory for studying the critical
 phenomena of the lattice scalar field theory with a global symmetry. 
 From a realistic point of view as well, it is significant  in connection
 with superfluid ${}^4$He 
\ref\OJ{T. Ohta and D. Jasnow, \PR{B20}, 139 (1979)}. 
This model has widely been studied  
\ref\LT{Y-H Li and S. Teitel, \PR{B40}, 9122 (1989)}
\ref\HM{M. Hasenbusch and S.Meyer, \PL{241}, 238 (1990)}
\ref\J{W. Janke, \PLA{148}, 306 (1990)}
\ref\GH{A. P. Gottlob,  M. Hasenbusch and S. Meyer,
\NPs{30}, 838 (1993) ; 
A. P. Gottlob and M. Hasenbusch, (unpublished)}
\ref\GZ{J.C. Le Guillou and J. Zinn-Justin, \PR{B21}, 3976 (1980)}
\ref\FMW{M. Ferer, M. A. Moore and M. Wortis, \PR{B8}, 5205 (1973)}
\ref\AHJ{J. Adler, C. Holm and W. Janke, (unpublished)
}
\ref\BCG{P. Butera, M. Comi and A. J. Guttman, \PR{B48},13987 (1993)}.
It is therefore suitable to choose this model and to compare with those analysis in order 
to check whether the chiral perturbation theory works.
 The advantage of taking the model of the dimension three ($d=3$) in the chiral perturbation theory
  is that finite size corrections to the partition function  are uniquely 
fixed only by the two (three)  low energy constants  in the $\epsilon$- ($p$-) expansion 
 not only to the leading order $1/L$ but also to the next  leading 1/$L^2$\HL. 
 They are the magnetization $\Sigma$ and the helicity modulus $\Upsilon$
(and an additional constant for the $p$-expansion)  in  infinite volume.
The helicity modulus is a quantity associated with the response to the distortion 
of the direction of the spontaneous symmetry breaking
\ref\FBJ{M.E. Fisher, M.N. Barber and D. Jasnow, \PR{A8},1111 (1973)}.
An increment of the free energy per volume due to the distortion is given by
 $\Delta f = \Upsilon (\Delta \epsilon)^2 /2$,
 where $\Delta \epsilon$ stands for the gradient of the twist in one spatial direction .
The equivalence between the helicity modulus  and the Goldstone boson decay constant $F$ is 
 shown in ref.\HL; $\Upsilon = F^2$.
 A numerical study of the finite size behavior of the system would then 
provide these constants and their associated  critical indices in a high accuracy. 
To the $d=4$ O(4) scalar model
\ref\HJJKLLN{A. Hasenfratz, K. Jansen, J. Jers\'ak, H.A. Kastrup, C.B. Lang, H. Leutwyler and 
T. Neuhaus, \NP{356}, 332 (1991)} 
\ref\GoL{M. G\"ockeler and H. Leutwyler, \PL{253}, 193 (1991) }
\ref\GJN{M. G\"ockeler, K. Jansen and T. Neuhaus, \PL{273}, 450 (1991)}
and 
the  $d=3$ classical Heisenberg model \ref\DHNN{I. Dimitrovi\'c, P. Hasenfratz, J. Nager 
and F. Niedermayer, \NP{350}, 893 (1991) } the chiral perturbation theory has successfully been  
 applied. 
\par
As stated above, each expansion has its own characteristic region, which manifests
 itself in the external source plane.
It is then  worthwhile to have a careful look at  where each of them is located and whether
 the two  overlaps or not. The authors of the paper \HJJKLLN\ have addressed this issue.
 We shall study this aspect  in  more detail. \par
 This paper is organized as follows. In section 2 we define the model and fix the notations.
We summarize the formulae of the $\epsilon$- and $p$- expansions for the magnetization, 
 the susceptibilities and the two point correlation functions. In section 3
we present the results of our numerical simulations based upon the cluster algorithm. 
 Section 4 is devoted to conclusions 
and discussion.

\newsec{The $XY$ model and  formulae  of the chiral perturbation theory  in $3$ dimensions}
The $O(2)$ linear $\sigma$ model  coupled with a constant external source $j$ 
 is defined  in the continuum euclidean space time as 
\eqn\lsmc{
\CL = {1 \over 2}( {\partial_\mu}\phi^i(x) )^2 +
      {1 \over 2}m^2( \phi^i(x) )^2 +
      \lambda ( ( \phi^i(x) )^2 )^2 -
       j\phi^0(x)
}
where $\phi(x)$ is a two component real vector ($i$= 0 and  1),  and $m$ and
 $\lambda$ stand for the  bare mass and
 the bare  quartic  coupling  constant, respectively.
 The external source has  only $i=0$ component.
To put the model on a lattice, one  rescales all the dimensionful quantities by
 a lattice constant $a$, and conventionally    uses three  
dimensionless parameters; the  quartic coupling parameter $\lambda_{lat}$,  the
hopping parameter $\kappa$ and the external source $J$. The lattice action is 
\eqn\lsml{
S =-\kappa \sum_{n,\mu,i} {\varphi_n^i \varphi_{n+\mu}^i } +
    \sum_{n,i} {\varphi_n^i \varphi_n^i } + 
    \lambda_{lat} \sum_{n} \Bigl(\sum_{i} \varphi_n^i \varphi_n^i - 1 \Bigr)^2 -
    J \sum_n \varphi^0_n
}
where $\varphi^i_n$ is the dimensionless field  $\varphi^i_n = (1/{\sqrt{\kappa}})\ \phi^i(na)
a^{d/2-1}$ 
sitting at site $n$ and $\mu$  is a unit vector of the  $\mu$ direction.
The lattice  parameters are related to those of continuum in such a  way as
$(ma)^2=(2-4\lambda_{lat})/\kappa-2d$, $\lambda a^{4-d}=\lambda_{lat}/\kappa^2$,
 $ja^{1+d/2}=J/ \sqrt{\kappa}$.
For $\lambda_{lat}=\infty$, the radial mode of 
$\varphi_n $ is frozen to unity, and the action  reduces to that of the  O(2) non-linear sigma
 model  or the $XY$ model. \par
Our aim  in the present  paper is to compute, in the $d=3$   $XY$ model, 
 two low energy constants in infinite volume, which  are the  
magnetization    $\Sigma$
\eqn\sig{
\lim_{j \to 0} \lim_{V \to \infty}\vev{\phi^0}_{j,V} = \Sigma
}
and the helicity modulus (or the Goldstone boson decay constant) $F$ appearing in the following 
equation 
\eqn\F{
\int dx\ e^{i p x} \vev{A_\mu(x) A_\nu(0)}_{j=0,V=\infty}
 = \delta^{\mu \nu} {i F^2 \over p^2} + \cdots
}
where $A_\mu = \phi^0 \partial_\mu\phi^1 - \phi^1 \partial_\mu\phi^0 $. The quantity 
$\vev{ \phi^0 }_{j,V}$ is the expectation value of $\phi^0$ in a three dimensional box with 
finite volume $V$  in the presence of the external source $j$. 
\par
In $d = 3 $, all the formulae
of the chiral perturbation theory are  basically 
fixed only by these  two low energy constants  to $O(1/L^2)$.
 It is   unlike the four dimensional models where
 a few additional   constants are necessary\HL. 
In the following,  we summarize   the formulae of the chiral perturbation theory
 necessary  to  our  analysis of finite size effects on the magnetization, the susceptibility 
and the two point functions. 
Detail about the formulae  is found  in ref.\HL\ for the $\epsilon$-expansion and 
 in  ref.\ref\GoLA{M. G\"ockeler and H. Leutwyler, \NP{361}, 392 (1991) }
 for the $p$-expansion.\par
In order for the chiral perturbation theory to be effective,  the Goldstone modes should be 
important   at long distances.  That is to say, finite size effects from the 
massive component with a mass $m_\sigma$  must be negligible
 compared to the Goldstone boson mass $m_\pi$, i.e., 
$m_\sigma / m_\pi \gg 1$.  Assuming the dominance of the Goldstone modes 
two manners  of expansions are possible depending upon what  quantities are fixed in
 the expansions\HL.
One is called $\epsilon$-expansion, in which 
the total magnetic energy $u_0 =\Sigma j L^d$  is fixed to $O(L^0)$, and another is 
$p$-expansion
 where $v=\Sigma j L^2 / F^2$ is fixed to $O(L^0)$. The latter gives   $j\sim L^{-2}$, while 
the former does  $j\sim 1/L^d$.   By using the Goldstone mass $m_{\pi}$, which is given
 by $m_{\pi}^2 =  \Sigma j/ F^2$, or the corresponding correlation length $\xi_{\pi}=1/m_{\pi}$,
 the domain is characterized by $m_{\pi}L \lsim 1 \quad (\xi_{\pi} \gsim L )$ for the 
 $\epsilon$-expansion and by $m_{\pi}L \gsim 1  \quad (\xi_{\pi} \lsim L)$ for the $p$-expansion.
Since  $m_{\pi}L =  \sqrt{\Sigma j } L / F   \lsim 1$ yields $ j \lsim F^2/\Sigma L^2 = {\rm const.}$ for a fixed 
$\kappa$ and $L$, 
one expects in  the $j$ space that the $\epsilon$- ($p$-) expansion holds  in the 
smaller (larger) $j$ domain. 
As one moves $\kappa$, each of the  regions  shifts.  In the vicinity of the critical point $\kappa_c$,
 $\Sigma$ and $F$ behave as $\Sigma \sim (\kappa - \kappa_c)^{\nu (1+\eta )/ 2}$ and
$F\sim (\kappa - \kappa_c)^{\nu / 2}$, respectively.
 One   hence expects that $m_{\pi}L = \sqrt{\Sigma j} 
L / F \approx 1$ leads to $j \sim F^2 / \Sigma \sim (\kappa - \kappa_c)^{\nu (1-\eta)}$ for
 a fixed $L$.
Therefore the domain for the
 $\epsilon$-expansion shrinks as $\kappa$ approaches $\kappa_c$ from above,
 since the index $\eta$, anomalous dimension of the field, is smaller than unity for 
the model in consideration.
It is one of our motivations to locate the region of 
 each expansion in the $j$ space.  
 It is also interesting to study 
 to what extent each region is  extended  beyond  the point where $m_{\pi}L \approx 1$. 
  This is another motivation of our work.
 In the present paper we shall look into these aspects in detail.
\par
All the formulae   in this section are of order $O(L^{-2})$.
Hereafter we call the direction  of the external source ($i=0$) the longitudinal direction
  and another direction  ($i$=1) the transverse one. They are denoted as $\|$
 and  $\bot$, respectively.
We work only with    finite volume  and finite $j$,
and  drop, for simplicity,   the suffices $j$ and $V$ from  $\vev{ \phi^0 }_{j,V}$.  
  As an expansion parameter we use $\alpha (=1/{F^2 L})$.\par

\subsec{$\epsilon$-expansion}
    The magnetization  in  the direction of an external  source is
\eqn\evev{
\vev{\phi^0}=\Sigma u [\rho_1 \eta + 2 \rho_2 \alpha^2]
,}
where $u = \rho_1 \Sigma j V$,  and $\rho_1$ and $\rho_2$ are  quantities depending  on
the shape of the box;
\eqn\xrho{\eqalign{&
\rho_1 = 1 + {1\over2}\beta_1\alpha + {1\over8}(\beta_1^2-2\beta_2)\alpha^2
\cr &
\rho_2 = {1\over4}\beta_2
.}}
For a three dimensional symmetric  box,   $\beta_1 =0.225785$ and $\beta_2 =0.010608$. 
The quantity $\eta $ in \evev\ is given by the  modified Bessel's functions $I_n(x)$ as
\eqn\xeta{
\eta = {1 \over u}{I_1(u) \over I_0(u)}
.}
Two point  correlation functions are defined as
\eqn\defcor{
G^{i j}(t) =
  {1 \over L^2}\sum_{\bf n} \vev{\phi^i({\bf n},t) \phi^j({\bf 0},0)}
}
where the summation is taken over in  spatial 2-volume  at fixed ``time" $t$ 
 ($0\leq~t<L$).
We consider two correlation functions; the longitudinal  and transverse ones;
\eqn\corf{
G_\|^\epsilon(\tau)    = G^{00}(t), \qquad   G_\bot^\epsilon(\tau)   = G^{11}(t) 
,}
where  $\tau=t/L$ and $0 \leq\tau < 1$.
In the $\epsilon$-expansion  each correlation function  takes the following form
\eqnn\el
\eqnn\et
$$\eqalignno{
G_\|^\epsilon(\tau)   &  \equiv
 a_\| + b_\| h_1(\tau) + c_\| h_2(\tau) + d_\| h_3(\tau)
,&\el \cr
G_\bot^\epsilon(\tau) & \equiv
 a_\bot + b_\bot h_1(\tau) + c_\bot h_2(\tau) + d_\bot h_3(\tau)
,&\et \cr
}$$
where coefficients $a_\|, b_\|, c_\|$  and $d_\|$  and their transverse counterparts are 
 expanded in powers of $\alpha$.  The $\tau$ dependence enters through the following   kinematic
 functions
\eqn\xxx{\eqalign{&
h_1(\tau)={1 \over 2}\Bigl( (\tau-{1 \over 2})^2 - {1 \over 12} \Bigr)
\cr &
h_2(\tau)={1 \over 24}\Bigl( \tau^2 (1-\tau)^2 - {1 \over 30} \Bigr)
\cr &
h_3(\tau)={h_1(\tau)}^2
         +{\sum_{\bf n}}^{'}{{ \cosh\bigl(q_{\bf n}(\tau-1/2)\bigr) }
                          \over
                          { 2 q_{\bf n} \sinh({q_n}/2) }
                         }
         -\beta_2
.}}
The  summation on  the r.h.s. of  $h_3(\tau)$ is taken over  all integers $(n_1,n_2)$
except for  ${\bf n}=(0,0)$, and $q_{\bf n}$ stands for 
$
q_{\bf n} = 2 \pi \sqrt{  n_1^2 + n_2^2 }
$.
The coefficients $a,b,c$ and $d$  for the longitudinal correlation function in \el\ are given by 
\eqn\xxx{\eqalign{&
a_\| = \Sigma^2 \left\{ \rho_1^2 (1-\eta) + 2 \rho_2 (1+2u^2\eta)\alpha^2 \right\}
\cr &
b_\| = \Sigma^2 \rho_1^2 \eta \alpha
\cr &
c_\| = \Sigma^2 ( 1 - 2 \eta ) \alpha^2
\cr &
d_\| = \Sigma^2 ( 1 - \eta ) \alpha^2 /2
,}}
while  the  transverse ones  in \et\ are given by 
\eqn\xxxx{\eqalign{&
a_\bot = \Sigma^2 \left\{ \rho_1^2 \eta + 2 \rho_2 \alpha^2 \right\}
\cr &
b_\bot = \Sigma^2 \rho_1^2 (1 - \eta ) \alpha
\cr &
c_\bot = \Sigma^2 \left( -1 + (2+u^2) \eta \right) \alpha^2
\cr &
d_\bot = \Sigma^2 \eta \alpha^2 /2
.}}
Note that $a, b, c$ and $d$ are of $O(1), O(\alpha), O(\alpha^2)$ and $O(\alpha^2)$,
respectively,  for both the cases.
We use these correlation functions to extract the low energy constants.
In addition to these,    we find that 
 the scalar product correlation function  is  also practically of much use;
\eqn\ev{\eqalign{
G_s^\epsilon(\tau) & =
   {1 \over L^2}\sum_{\bf n} \vev{\phi({\bf n},L\tau) \cdot \phi({\bf 0},0)}
\cr
          & = G^\epsilon_\|(\tau) + G^\epsilon_\bot(\tau)
\cr
          & = a_s + b_s h_1(\tau) + c_s h_2(\tau) + d_s h_3(\tau)
,
}}
where the coefficients $a_s$ etc. are given by
\eqn\gs{\eqalign{&
a_s = \Sigma^2 \left\{ \rho_1^2  + 4 \rho_2 (1+u^2\eta)\alpha^2 \right\}
\cr &
b_s = \Sigma^2 \rho_1^2  \alpha
\cr &
c_s = \Sigma^2 u^2 \eta \alpha^2
\cr &
d_s = \Sigma^2  \alpha^2 /2
.}}
The longitudinal susceptibility is calculated from it's correlation function 
\eqn\defsus{
\chi_\|   = \sum_n \left( \vev{\phi^0(n) \phi^0(0)} - \vev{\phi^0}^2 \right)
,}
and it takes the form 
\eqn\esus{
\chi_\|    =  \Sigma^2 V \left[ \rho_1^2\{(1-\eta) - u^2 \eta^2\} + 2 \rho_2 \alpha^2 \right]
.}
\subsec{$p$-expansion}
In the $p$-expansion to  $O(L^{-2})$
 a  low energy constant $k_0$ gets involved in addition to  $\Sigma$ and $F$
for  the magnetization and the  susceptibilities.\par
The magnetization  reads 
\eqnn\pvev
$$\eqalignno{
\vev{\phi^0} = & \Sigma\left\{ 1  + {1 \over 8}\alpha \zeta^{1/2} + k_0 \alpha^2 \zeta -
    {1 \over 2}\alpha\bigl[1 + {3 \over{16\pi}}\alpha \zeta^{1/2}\bigr]g_1(\zeta_p) \right.\cr 
    &\qquad\, +
    \left.{1 \over 8}\alpha^2 g_2(\zeta)\bigl[g_1(\zeta) - \zeta g_2(\zeta)\bigr]
    \right\}
,&\pvev \cr
}$$
where $ \zeta=(m_{\pi} L)^2 $  (  $ m_{\pi}^2=\Sigma j / F^2 $) , and $\zeta_p$ is
\eqn\xxip{
\zeta_p 
      = \zeta \bigl( 1 + {1 \over {8\pi}}\alpha\zeta^{1/2} \bigr)
.}
The function $g_n(\zeta)  (n=0,  1, 2, ...)$ are defined by
\eqn\g{
g_0(\zeta) = - \ln \zeta - {\zeta^{3/2} \over {4\pi}} - 
             \sum_{n=0}^\infty {\beta_n \over n!}\zeta^n
}
\eqn\g{
g_{n+1}(\zeta) = - {d \over d\zeta}\ g_n(\zeta)
.}
\par 
The correlation functions  defined in \corf\  and \ev\  take the forms 
\eqnn\pl
\eqnn\pt
$$\eqalignno{
G_\|^p(\tau)   & = {\vev{\phi^0}}^2 + {1 \over 2}\Sigma^2 \alpha^2 \bar{h}_2(\tau)
& \pl \cr
G_\bot^p(\tau) & = Z_v \alpha \bar{h}_1(\tau)  
,& \pt \cr
}$$
where $Z_v$ is the wave function renormalization constant in a finite box with the form
\eqn\zv{
Z_v = \Sigma^2 \left[ 1 +{1 \over 4\pi}\alpha\zeta^{1/2} - g_1(\zeta)\alpha \right]
.}
The kinematic functions  $\bar{h}_1(\tau)$ and $ \bar{h}_2(\tau)$ involved in the $p$-expansion 
 are 
$$\eqalign{
\bar{h}_1(\tau) & = \bar{h}(\tau,\zeta_v)
\cr 
\bar{h}_2(\tau) & = \sum_{\bf n} \bar{h}^2(\tau,\zeta + 4\pi^2|{\bf n}|^2)
}$$
 where $\bar{h}$ on the r.h.s. is given by 
$$
\bar{h}(\tau,\zeta)  = {1 \over {2 \sqrt\zeta}}
               {\cosh{\left(\sqrt\zeta(\tau-1/2)\right)} \over \sinh{\sqrt\zeta/2}}
$$
and $\zeta_v $ is 
\eqn\zetav{
\zeta_v = \zeta_p \left[1 - {1 \over 2}g_1(\zeta) \alpha \right]
.}
The transverse correlation function has a  normal form of massive scalar propagator to
  $O( \alpha )$,
 while the connected part of the  longitudinal one \pl\ is $O(\alpha^2)$.
We  also consider the scalar product correlation function
\eqn\pv{
G_s^p(\tau) = G_\|^p(\tau) + G_\bot^p(\tau)
.}
As in the $\epsilon$-expansion, $G_s^p(\tau)$ is of practical use.\par
%
The susceptibility in \defsus\ in terms of $p$-expansion reads 
\eqnn\psus
$$\eqalignno{
\chi_\| & = \Sigma V \alpha \left\{ {1 \over {16\pi}}\alpha \zeta^{1/2} + k_0 \alpha^2 \right.\cr
        & \qquad\qquad -
          {1 \over 2}\alpha\Bigl[-g_2(\zeta_p)-{3 \over {32\pi}}\alpha\zeta^{1/2}
                     \bigl( g_1(\zeta_p)+2(1+\zeta)g_2(\zeta_p) \bigr) \Bigr] \cr
        & \qquad\qquad +
           \left. {1 \over 4}\alpha^2\bigl[-2g_1 g_2 + \zeta( {g_2}^2+g_1 g_3 ) \bigr]
           \right\}
.&\psus \cr
}$$
\newsec{Numerical simulations}
We employ, to generate  configurations, the Wolff's single cluster algorithm
\ref\W{U. Wolff, \PRL{62}, 361 (1989) }
 and its modification in the case of presence of an external source\DHNN.
The lattice shape which we take  is a symmetric box with  a volume $V=L^3$,
 and  the size of  the box ranges from $L=32$ to $L=64$.
The $d=3$ dimensional $XY$   model is known to have  a second order phase transition 
at $\kappa_c=0.45420(2)$\GH.
 We calculate  at  $\kappa$ larger than $\kappa_c$, and choose  two
$\kappa$ values $\kappa$=0.462 and 0.47 for main calculations of the
low energy constants, and several more for obtaining the critical indices. 
The magnitudes of the external source changes   from $J=0.0$ to 
 $J=5.0 \times 10^{-4}$. In terms of continuum external source, which depends on $\kappa$, 
 $j$ ranges from 0 to 7.29 $\times 10^{-4}$ for, say,  $\kappa = 0.47$.
We make measurements at each 5  updatings with 
typical number of configurations for  measurements ranging from about 20,000 to 40,000.
The statistical errors are estimated by using the  blocking  and the bootstrap methods. 
We use the program system  SALS 
\ref\SALS{T.Nakagawa and Y.Oyanagi,
`Program System SALS for Nonlinear Least-Squares Fitting in Experimental Sciences',
in Recent  Developmetnt in Statistical Infernce and Data Analysis,
North-Holland(1980)}
for the least square fitting of the data to the formulae.\par
We extracted $\Sigma$ and $F$ from three types of correlation functions in each of the 
$\epsilon$- and $p$- expansions.
We looked at the behaviors of each of the extracted values 
by varying $J$. We also observed the magnetization and the susceptibility, and compared the results 
with the formulae of the chiral perturbation theory.
Main results are in order.
\item{(1)} We found, as expected from the arguments in the Sec. 2,
 that the \hbox{$\epsilon$-expansion} provides good results in 
the small $J$ region, while the \hbox{$p$-expansion} does so in the larger $J$ region.
 We identified where the   region of the validity 
for  each of the two manners of  expansions is located.
\item{(2)}The scalar product correlation functions provide stable results  for the two
 constants. 
\item{(3)}Upon  varying the size of the lattice, which is sensitive to the region of the validity,
 we observed a significant change of the behavior for the \hbox{$p$-expansion}.  It is 
consistent with the arguments concerning $m_\pi$ and $m_\sigma$.\par
\item{(4)}The magnetization 
 shows a clear crossover behavior between the two expansions as a function of $J$.
 The longitudinal susceptibility turns out too poor to address the region of 
 the validity.
\item{(5)} We calculated the critical indices associated with the two constants.  The results are 
consistent with other references.
\par
In the following we shall discuss the results in detail.
\subsec{Correlation functions}
In order to   extract   $\Sigma(\kappa)$ and $F(\kappa)$ from the correlation functions,
we fit to the  formulae \el, \et\  and \ev\
 for the $\epsilon$-expansion  and to \pl, \pt\ and \pv\ for the \hbox{$p$-expansion}
 for each  $j$  at fixed $\kappa$.
We use hereafter the notation  $\Sigma_{(\epsilon , p ) , ( || , \bot,s)}$
 for the value of $\Sigma(\kappa)$ obtained from
each type of correlation functions in each expansions.  
Similar notations are used also  for  $F$.
All the extracted values  of $\Sigma$ and $F$ are listed in Table I -- V.
In what follows we discuss the stability of the extracted results and the region of the  validity
 of each of the expansions by comparing the values in  Tables. If the chiral perturbation 
theory works, the extracted values should be independent of $J$, 
and all of the values $\Sigma_{(\epsilon , p ) , ( || , \bot,s)}$
 ($F_{(\epsilon , p ) , ( || , \bot,s)}$) should agree.\par
\bigskip
\leftline{{\bf  Results of $L=32$ lattice ($\kappa=0.47$) }}\par
In Figs. 1 and 2 we show the behaviors of the  obtained 
 $\Sigma(\kappa)$ and $F(\kappa)$ v.s. the magnitude of the external source $J$ 
at  $L=32$ for $\kappa$=0.47.
In Fig. 1 we see  that the $\epsilon$-expansion  provides the consistent  result for $\Sigma$
 within  errors 
 for two  types of correlation functions, i.e., 
 longitudinal one $G_\|^\epsilon(\tau)$ and  transverse one $G_\bot^\epsilon(\tau)$,   
in the whole  $J$ range in consideration. 
The $p$-expansion, on the other hand, is out of validity in the smaller $J$ region, $J \lsim 2.0 
\times 10^{-4} $ (Fig. 1).  At  larger $J$ values, the transverse  one  
$\Sigma_{p,\bot}$ agrees with $\Sigma$ of the $\epsilon$-expansion.  
  We hence observe that 
 an overlap of the two expansions takes place at $J \gsim 2.0\times 10^{-4}$. 
 \par
The result of the  longitudinal $p$-expansion is not included in the figure.
Fittings  according  to this type of expansion turn out invalid in the whole $J$
 range.
This may be  due to  the fact that  $\tau$ dependence appears
 only at $O(\alpha^2)$ unlike the all other correlation  functions of   $O(\alpha)$, 
 and that   the data would be  too   noisy to extract the $\Sigma$.
This feature applies also to $F$ and to different volumes.  We then drop the result of
the  longitudinal $p$-expansion  throughout the paper. \par
Fig. 2 shows $F$ v.s. $J$.
    Both  of  $F_{\epsilon,  || }$ and   $F_{\epsilon,  \bot}$ 
 agree  within errors  
 in the small $J$ region ($J \lsim  2.0 \times 10^{-4}$),
  while  $F_{\epsilon,  || }$ and   $F_{\epsilon,  \bot}$   split from each other 
 for  larger $J$ ($J \gsim  2.0 \times 10^{-4}$).  
This    indicates that    
 the boundary of the two  domains where the $\epsilon$-expansion is valid and invalid
is located at $J \approx  2.0 \times 10^{-4}$.
 Fig. 2 also  shows the behaviors of $F$ for the \hbox{$p$-expansion}.
Similarly to $\Sigma$, the $p$-expansion does not give a reasonable result in the smaller $J$
 region, but agree with $F$ of $\epsilon$-expansion at 
$ J \gsim 2.0 \sim 3.0 \times 10^{-4}$.
\par
The lowest value of the overlapping region for $\Sigma$ and $F$
 is consistent with what one naively  expects 
 from the  argument concerning the Goldstone mass (Sec. 2);
 the two expansions coincide at  a point $J$ which is determined by the relation 
$M_{\pi}L=\sqrt{\Sigma j } L /F \approx 1$.
Taking  the values   $\Sigma(0.47)=0.2581(1)$,  $F(0.47)=0.286(1)$ and $L=32$
for $\kappa=0.47$,  it  yields    $ 2.1\times 10^{-4}$,   
which is in good agreement with the observed value. 
Note that the errors of $F$ are about five times larger than those of $\Sigma$
 at whole $\kappa$'s and $J$'s in consideration.   
\par
\bigskip
\leftline{{\bf  Results of $L=32$ lattice ($\kappa=0.462$) }}\par
Similar behaviors are seen  for the data of $\kappa=0.462$ and $L=32$ (Fig. 3). 
 However, 
 $\Sigma_{\epsilon, \|}$ and   $\Sigma_{\epsilon, \bot }$  agree only at small $J$ values 
($J \lsim 2.0 \times 10^{-4}$)  compared to $\kappa=0.47$ (Fig.~\ 1). 
This is in fact what is expected  from the condition  giving the region  of the expansion 
  $M_{\pi}L \lsim 1$, since the region of the $\epsilon$-expansion shrinks according to 
$j\lsim(\kappa-~\kappa_c)^{\nu(1-\eta)}$
 as $\kappa$ approaches $\kappa_c(\kappa_c = 0.4546)$ as stated in Sec. 2.
We see that this  behavior becomes  even clearer for $F$ (see Table II).
\par
Fig. 3  also shows  the behaviors of the $p$-expansion.
  The value    $\Sigma_{ p ,  \bot}$ 
 seems reasonable only in the limited intermediate region around 
$J \approx 2.0 \times 10^{-4}$.  In larger $J$ region the \hbox{$p$-expansion} becomes  
invalid unlike  the case of $\kappa=0.47$.
The reason for that  comes from the mass   $m_\sigma$ of  the massive particle;
 as stated in Sec. 2, $m_\sigma$ must be large  enough $ m_\sigma / m_\pi \gg 1$.
In the  case of $\kappa=0.47$ we obtained   $m_\sigma \approx 0.41$, which gives the ratio 
$m_\sigma / m_\pi$ ranging  from 20.5 to 10.1 as $J$ changes from $1.0 \times 10^{-4}$
 to $5.0 \times 10^{-4}$.
$m_\sigma$ is then heavy enough for the chiral perturbation to apply at $\kappa=0.47$.\par
For $\kappa=0.462$, on the other hand, we obtained smaller value $m_\sigma = 0.3$,
 which yields $m_\sigma / m_\pi$ ranging  from 12.5 to 5.5 in the same $J$ range.
The authors of  the paper \HJJKLLN\ 
took  $m_\sigma / m_\pi \gsim 10$ as a criterion.
If we employ the same value, it provides an upper bound of $J$, $J \lsim 1.5\times 10^{-4}$,
on the region of the validity of the chiral perturbation theory.
Combining the condition  $m_\pi L \gsim 1$, which gives a lower bound of $J$, 
 $J \gsim 1.66 \times 10^{-4} $, on the
region of the $p$-expansion, one observes that there is no region for the
 validity of the  $p$-expansion.
If  $m_\sigma L\gsim 7.5 $  is adopted instead of 10, however,  
the upper bound shifts to $J=2.7 \times 10^{-4}$.
  Although these bounds should not be taken strictly, it indicates that the $p$-expansion
 is expected to apply to  no or only  the narrow $J$ region around $J=1.6 \sim 2.5 \times
 10^{-4}$.
In other words, the ``window'' for the $p$-expansion is almost closed or  open only slightly. 
The results of our fittings are in  good agreement with   this inference (Fig. 3).
\par
\bigskip
\leftline{{\bf Scalar product correlation function }}\par
Apart from the longitudinal and transverse correlation functions,
 we made also use of   the scalar product correlation functions \ev\ and \pv.
During our fitting procedure we came to notice that the scalar product correlation functions
 are good correlators which provide   stable  fitting results. In particular, the 
$\epsilon$-expansion works very well.\par 
In Fig. 4 we  show the results of $\Sigma$ for both of the $\epsilon$- and $p$-expansions
 at $L=32$  for $\kappa=0.47$ and 0.462.
For both $\kappa$ values $\Sigma_{\epsilon,s}$ looks  almost independent of $J$ in the 
whole $J$ range, and  therefore it   provides  reliable  result.
This is the reason why we indicated the value $\Sigma_{\epsilon,s}$ in Figs. 1 and 3 by 
 the  arrow as a reference.
In the region of validity of the $\epsilon$-expansion at each $\kappa$, $\Sigma_{\epsilon,s}$
 is  consistent with $\Sigma_{\epsilon,\|}$ and $\Sigma_{\epsilon,\bot}$.
Compared to the other two, $\Sigma_{\epsilon,s}$ is  much stable. 
\par
The value $\Sigma_{p,s}$, on the other hand, vary as $J$ moves.
For  $\kappa=0.47$  it   monotonically approaches $\Sigma_{\epsilon,s}$
  as   $J$  increases, and gives consistent result at 
  $J \gsim 3.0 \times 10^{-4}$. For $\kappa=0.462$, on the other hand, $\Sigma_{p  , s}$ 
 never agrees with  $\Sigma_{\epsilon  , s}$ even in the larger $J$ region. 
  The scalar product correlation functions    supports more clearly, than do the longitudinal
 and the transverse correlation functions, 
  the inference based upon  $m_\sigma$ and $m_\pi$ about the location of the boundary  
 between the two expansions. 
Similar behaviors are seen also for $F$ as shown in  Fig. 5.\par
\bigskip
\leftline{{\bf Volume dependence }}\par
    Let us turn to   the volume dependence.
The extracted values of $\Sigma$ and $F$ for $L=48$ are listed in Tables III and IV .
The average values for $L=48$ are consistent with those for $L=32$ in the regions of
 the validity of each of the expansions.\par
When one increases  the volume, $J$ value corresponding to the condition
 $m_\pi L \approx 1$
 decreases  as $J \propto 1/L^2$, while the ratio $m_\sigma / m_\pi$ is independent of $L$.
It is therefore expected that the ``window'' for the $p$-expansion becomes wider as $L$
 increases. It is then interesting  if one actually sees this behavior by  simulations. 
 For $L=32$ at  $\kappa=0.462$, we have seen that  the ``window''  is almost closed.   
Fig. 6  shows $\Sigma_{\epsilon, s}$ and $\Sigma_{p , s}$ 
 for $L=48$  at  $\kappa=0.462$ obtained from the scalar correlation functions.  
The value $\Sigma_{p  , s}$  agrees with $\Sigma_{\epsilon  , s}$ only around  $J=2.0 \times 10^{-4}$,
 which  is regarded as an evidence of  slightly opening  ''window''.  It is in good contrast
 with  the behavior of $\Sigma(\kappa=0.462)$  in  Fig. 4.  
 Similar behavior  is found also for $F$ as shown in Figs. 5 and 7.  \par
%
In the absence of $J$ we calculated $\Sigma$ and $F$ at $L=64$, and compared them
 with those of $L=32$ and $48$.  As seen in Table V the results of $\Sigma_{\epsilon, s}$ 
 and $F_{\epsilon, s}$ are consistent within errors.\par

\subsec{Magnetization and susceptibility}
We calculate the magnetization for the $\epsilon$-expansion
 \evev\ for the $p$-expansion \pvev.
Rather than fitting  \evev\ and \esus\ to the data to extract $\Sigma$ and $F$,
 we put $\Sigma$ and $F$  into eqs. \evev\ and \pvev\ and
 compare them  with those of direct measurements of $\vev{\varphi^0}$. 
Here we use the values of $\Sigma$ and $F$  obtained
 from the correlation functions at each $\kappa$,
 $\kappa=0.462$ and $\kappa=0.47$, in the previous subsection.
In the $p$-expansion, an additional low energy constant $k_0$ is involved 
 as mentioned before.
To determine $k_0$, we match the  curve \pvev\ to the data  at some $J$.
\par
Fig. 8 shows the results for $\kappa=0.462$ and $0.47$ at $L=32$.
The  constant $k_0$ is fixed at  $J=2.0 \times 10^{-4}$ for $\kappa=0.462$
 and  at $J=3.0 \times 10^{-4}$ for $\kappa=0.47$.
Solid line in the figure shows the curve \evev\ and dotted line is \pvev.
We see a good agreement and observe a clear crossover between the $\epsilon$- and 
the $p$-expansions.
The location of the  crossover  between both the  expansions  is consistent with 
 the prediction $m_\pi L \approx  1$.\par 
The results of $L=48$ are shown in Table VI. 
The values obtained from the $p$-expansion formula \pvev\ are $\vev{\varphi^0}=0.201$ at 
$J=0.5 \times 10^{-4}$,  $0.2934$ at $J=2.0\times 10^{-4}$
 and $0.3083$ at $J=3.0\times 10^{-4}$ for $\kappa=0.462$.
The two of them ($J=0.5 \times 10^{-4}$ and $2.0\times 10^{-4}$) are consistent with the
numerical result within errors.
Another one ( $J=3.0\times 10^{-4}$ ) seems to deviate slightly from the data. 
For $\kappa=0.47$, the values of $\vev{\varphi^0}$ read $0.279$
 at $J=0.5\times 10^{-4}$,
 $0.3357$ at $J=1.0\times 10^{-4}$ 
and $0.3776$ at $J=3.0\times 10^{-4}$, respectively.
In the similar manner to $\kappa=0.462$ only the two points ($J=0.5 \times 10^{-4}$ and
$1.0 \times 10^{-4}$) out of three are in good agreement with the data within errors.
In any case,   theoretical predictions are not inconsistent with the measurements. 
\par
Let us turn to  the longitudinal susceptibility $\chi_{||}$.
Fig. 9 shows the $J$ dependence of $\chi_{||}$ 
 for $L=32$ at $\kappa=0.462$ and $0.47$.
The solid (dotted) lines  correspond to  \esus (\psus). 
For each of the fixed $\kappa$, $\chi_{||}$ shows a crossover at $J \approx 2.0\times 10^{-4}$
 if the $\epsilon$- ($p$-) expansion is valid in the smaller (larger) $J$ region.
In the actual measurements  we are not able to  observe such a clear  crossover. 
\par
 As to the relation between  the lines for the  two
  $\kappa$'s we can read off qualitative difference  as follows.
For smaller $J$ ($J \lsim 2.0 \times 10^{-4}$),  
the values of $\chi_{||}(\kappa=0.462)$ 
are lower  than  $\chi_{||}(\kappa=0.47)$, while 
in larger $J$ region 
$\chi_{||}(\kappa=0.462)$  becomes higher than $\chi_{||}(\kappa=0.47)$.
This  property  is  successfully  seen in the direct measurements.
Quantitatively, however, there are some inconsistencies between the theoretical
 predictions and the numerical data. 
\par
The $J$ dependence of $\chi_{||}$ in the case of $L=48$ is  shown in Table VII.
The situation is the same as in  the case of $L=32$. 
\par
 
\subsec{Critical index}
The critical indices $\beta$ and $\nu$ are defined by
\eqn\crid{\eqalign{&
\Sigma(\kappa)  \sim ( \kappa - \kappa_c )^\beta \cr &
F(\kappa)  \sim ( \kappa - \kappa_c )^{\nu/2} 
}}
in  3 dimensions\HL.
To extract $\beta$ and $\nu$ we make use of  the values of $\Sigma$ and $F$ in infinite volume
obtained  from the previous  analysis.  We fit to \crid\  the data at  five $\kappa$ 
points ($\kappa=0.46, 0.462, 0.464, 0.466, 0.47$).
The results are
\eqn\kappac{
\kappa_c =  0.4546(5) 
}
for the critical coupling
\eqn\crid{\eqalign{&
\beta    =  0.321(17) \cr &
\nu      =  0.66(6)   
}}
for the indices.
They are consistent  with other references\LT\J\GH\GZ\FMW\AHJ\BCG.
\par
\newsec{Conclusions and discussion}
 We applied the chiral perturbation theory \`a la Hasenfratz and Leutwyler 
 to the $d=3$ $XY$ model in order to calculate the two low energy constants.  
They are the magnetization $\Sigma$ and
 the helicity modulus $F$ (or Goldstone boson coupling) in infinite volume.  
 In the theory, two manners of the expansions are involved. One is  the
$\epsilon$-expansion, which is valid in the region where $m_\pi L \lsim 1$, 
 and another is  the $p$-expansion, where $m_\pi L \gsim 1$.
On $L=32$, $48$  and $64$ lattices, we fitted the formulae of the  correlation functions 
to the Monte Carlo data. 
All the values of $\Sigma$ and $F$ extracted in each of the regions of the validity 
are  consistent and volume independent 
 within errors.\par
  We are also particularly concerned with the region of the validity of the two expansions.
As far as the two $\kappa$ values  are concerned, the lower boundary of the $p$-expansion
  is  basically located at   the region where $m_\pi L \approx 1$ holds. On the other hand,
the $\epsilon$-expansion stretches, for some cases, beyond   the point $M_\pi L \approx 1$
 to some extent than  expected.  The similar behavior  was also observed in ref.\HJJKLLN.
Apart from the condition for $m_\pi L $, the mass  $m_\sigma$ of the massive particle  
 puts a constraint on the validity of the chiral perturbation theory.
As expected from these conditions, we have observed the significant difference of  
 the behaviors of  $\Sigma$ and $F$ v.s. the external source $J$ for the two different
 $\kappa$  values. \par
We found that the scalar product correlation functions are better correlators than 
the longitudinal and transverse correlation functions. Particularly the $\epsilon$-expansion 
provides  quite stable estimations of $\Sigma$ and $F$. By use of it the overlapping of the
validity of  both the expansions is  clearly seen. The reasons for the  stability are in order. 
 The coefficient $b_s = \Sigma^2 \rho_1^2 \alpha$ in \gs\ is  independent of $j$, 
  unlike  the longitudinal $b_\|$ in \xxx\ and the transverse counterparts $b_\bot$ in \xxxx, in which 
$j$ dependence appears through $u=\rho_1 \Sigma j V$.
To $O(\alpha)$, therefore, no  response of $G^\epsilon_s$ to the variation of $j$ 
 appears. It is then expected that $\Sigma_{\epsilon, s}$ and $F_{\epsilon, s}$ are extracted 
 independently  of $j$.  This is one of the reasons for the stability
\footnote{*}{The fact that $b_s$ is independent of $j$ to $O(\alpha)$ applies
only to $N=2$ for the  $O(N)$ model.}. 
However for   $J \gsim 3.0 \times 10^{-4}$,  $\Sigma$ and $F$ 
  deviate from their stable  values, i.e., in this $J$ region the contribution from $O(\alpha^2)$
 would be  responsible for the stability.
In order to have a look at its  effect, we compared the 
 two ways of fitting of  $G^\epsilon_s$ using the formulae to $ O(\alpha)$ and to  $O(\alpha^2)$.
We found that  in the  small $j$ region $(J \lsim 3.0 \times 10^{-4})$ the results
 of the two fittings 
 are in agreement with each other  within errors, while in the larger region the significant discrepancy appears.
(The relative difference of the  two fits is
 approximately $4 \%$ for  $F(\kappa=0.47)$ at $J=5.0 \times 10^{-4}$)
One then sees that  the correction of $O(\alpha^2)$ plays a significant 
role to the stability in the larger $J$ region.
\par 
 The magnetization $\vev{\varphi^0}$ and the susceptibility $\chi_{||}$
  are fitted by use of 
the values of  $\Sigma$ and $F$ obtained by the correlation functions.
 The consistency was checked. The magnetization is well fitted by the $\epsilon$-expansion 
in the smaller $J$ region and by the $p$-expansion in the larger one.  A crossover  between 
 both the  expansions in the $J$ space was observed, and 
 its    location   is consistent with  the prediction $m_\pi L \approx  1$. 
 The susceptibility turns out too poor in  precision to address the issue about the validity 
of the expansions.\par
We calculated the critical indices associated with $\Sigma$ and $F$, and obtained 
 the values consistent with other references.
  However we have not reached the precision 
 as high as that of ref.'s \J\ and \GH, which is based upon
 the phenomenological renormalization group.
This may be  due to the fact that the chiral perturbation theory becomes hard to apply as it is. 
Because, as one approaches the critical point,  the mass $m_\sigma$ of the massive particle 
 gets smaller and the ratio
 $m_\sigma/m_\pi$ accordingly becomes small. In addition, the expansion parameter
 $\alpha = 1/F^2 L $ becomes large.
\par
A few words for the values of the constant $F$.
We compared the value of $\Upsilon$ calculated from $F$ with the one in the 
available references.  Our values\footnote{**}{Note that $F^2/\kappa$ corresponds to the helicity
 modulus in their  literatures.}
 $\Upsilon=F^2=0.0515(14)$ 
 at $\kappa=0.462$
 and $0.042(2)$ at $\kappa=0.460$ in infinite volume
 come slightly below  the curves in the figures of refs. \LT\ and \J, in which the lattice size
 $L$ is at most 48. 
 It is reasonable since in their data the values monotonically decrease 
as $L$ grows.

\eject
\bigskip
\centerline{\bf Acknowledgment}
We thank  M. Imachi for  a careful reading of the manuscript.
We are grateful to  M.Imachi and T. Kashiwa for encouragement and useful discussions, and 
 to the members of the high energy theory groups in Kyushu University and 
 Saga University for discussions.
The numerical simulation was performed on FACOM M-1800/20 at RCNP, Osaka university. 
\listrefs
 
\centerline{\bf Table Captions}
\smallskip
\item{Table  I}$\Sigma$ and $F$ for $\kappa=0.47$ and $L=32$.
The symbol *** indicates the point where the  $p$-expansion is beyond the applicability.
The symbol \#\#\# shows that we failed  to fit the formulae.
The same symbols  are used in the other tables.
\item{Table  II}$\Sigma$ and $F$ for $\kappa=0.462$ and $L=32$.
\item{Table  III}$\Sigma$ and $F$ for $\kappa=0.47$ and $L=48$.
\item{Table  IV}$\Sigma$ and $F$ for $\kappa=0.462$ and $L=48$.
\item{Table  V}$\Sigma$ and $F$ for $L=64$ and $J=0.0$.
\item{Table  VI}$\vev{\varphi^0}$ v.s. $J$ for $L=32$ and $L=48$.
By the symbol \%\%\% it is meant that  we have no data available.
\item{Table  VII} $\chi_{||}$ v.s. $J$ for $L=32$ and $L=48$.

\vfill\eject
\centerline{\bf Figure Captions}
\smallskip
\item{Fig. 1}{$\Sigma $ v.s. $J$ for $\kappa=0.47$ and $L=32$.
Circle indicates $\Sigma_{\epsilon, ||}$, square is $\Sigma_{\epsilon, \bot}$,  
and triangle corresponds to $\Sigma_{p, \bot}$. For small $J$ there is a significant difference
 between the $\epsilon$- and $p$-expansions.
The arrow shows the mean value of $\Sigma_{\epsilon,s}$, as a reference, which is
 calculated from $\Sigma_{\epsilon,s}$ at seven $J$ points in Table I.
The value $\Sigma_{\epsilon,s}$ is regarded as  the best estimate of $\Sigma$.
Detail about $\Sigma_{\epsilon,s}$ is found in the latter part in this section.
}
\item{Fig. 2}{$F $ v.s. $J$ for $\kappa=0.47$ and $L=32$.
Circle indicates $F_{\epsilon, ||}$, square is $F_{\epsilon, \bot}$,  
and triangle corresponds to $F_{p, \bot}$.
The location  of the arrow shows the mean value of $F_{\epsilon,s}$, as a reference,
 which is  estimated in the same manner as in Fig. 1 from Table I.
It is also the best estimate of $F$.
}
\item{Fig. 3}{$\Sigma$ v.s. $J$ for $\kappa=0.462$ and $L=32$.
Circle indicates $\Sigma_{\epsilon, ||}$, square is $\Sigma_{\epsilon, \bot}$,  
and triangle corresponds to $\Sigma_{p, \bot}$. For $J \gsim 2.0 \times 10^{-4}$ 
the $\epsilon$-expansion appears out of validity.
The arrow  shows the mean value of $\Sigma_{\epsilon,s}$.
}
\item{Fig. 4}{$\Sigma $ v.s. $J$ for $L=32$. All of the values $\Sigma$ are determined
 by the scalar product  correlation functions.
$\Sigma_{\epsilon,s}$ is almost independent of $J$ for both the $\kappa$ values.
For   $\kappa=0.47$  $\Sigma_{p,s}$ (square) agrees with $\Sigma_{\epsilon,s}$ (circle) 
 at $J \gsim 3.0 \times 10^{-4}$, while for  $\kappa=0.462$ there is a significant 
 difference between $\Sigma_{\epsilon,s}$ (filled circle) and $\Sigma_{p,s}$ (filled square).
The statistical errors of $\Sigma_{\epsilon,s}$ lie within the symbols. }
\item{Fig. 5}{$F $ v.s. $J$ for $L=32$. All of the values $F$ are
 determined by the scalar product  correlation functions.
 Similar behavior to $\Sigma$ in Fig. 4 is observed.  The same symbols  as  those in Fig. 4 
 are used.
}
\item{Fig. 6}{$\Sigma$ determined by the scalar product correlation functions at $L=48$
 and for  $\kappa=0.462$.  $\Sigma_{p,s}$ (square) agrees with $\Sigma_{\epsilon,s}$ 
 (circle) at $J=2.0 \times 10^{-4}$.  As seen in Table IV,
 the $p$-expansion does not provide reasonable value 
 for $\Sigma_{p,s}$ at $J=0.5\times 10^{-4}$, i.e, out of validity.
The arrow shows the mean value of $\Sigma_{\epsilon,s}$.
}
\item{Fig. 7}{$F $ v.s. $J$ at $L=48$ and  for $\kappa=0.462$. 
$F_{p,s}$ (square) agrees with $F_{\epsilon,s}$  (circle) at $J=2.0 \times 10^{-4}$.
The arrow indicates the mean value of $F_{\epsilon,s}$ 
}
\item{Fig. 8}{Magnetization $\vev{\varphi^0}$ v.s. $J$ for $L=32$.
Normalization of the data is done at $J=3.0 \times 10^{-4}$ for $\kappa=0.47$
 and at $J=2.0 \times 10^{-4}$ for $\kappa=0.462$.
The solid lines show \evev ($\epsilon$-expansion), and the dotted lines correspond to 
\pvev ($p$-expansion).  A clear crossover  between the two expansions is observed for both
   $\kappa=0.47$ (square) and  $\kappa=0.462$ (circle).
}
\item{Fig. 9}{Longitudinal susceptibility $\chi_{||}$  v.s. $J$ for $L=32$.
The solid lines show \esus\ ($\epsilon$-expansion), and the dotted lines correspond to 
\psus\ ($p$-expansion).
}
\vfill\eject

\def\el{\epsilon,||}
\def\et{\epsilon,\bot}
\def\es{\epsilon,s}
\def\pl{p,||}
\def\pt{p,\bot}
\def\ps{p,s}

\font\cour=cmtt10 scaled\magstep2

$$\vbox{
\halign{\hfil#\hfil&&\quad\hfil#\cr
$\Sigma$ \cr
\noalign{\smallskip\hrule\smallskip}
 $J(\times 10^{-4})$ & $\el$ & $\et$ & $\es$ & $\pt$ & $\ps$ \cr
\noalign{\smallskip\hrule\smallskip\hrule\smallskip}
$0.0$&$   0.2590(8)$&$   0.2569(8)$&$   0.2579(3)$&    ***&    ***\cr
$0.5$&$   0.2577(8)$&$   0.2576(9)$&$   0.2573(4)$&    \#\#\#&$   0.3780(5)$\cr
$1.0$&$   0.2567(10)$&$  0.2603(17)$&$   0.2577(3)$&$    0.272(2)$&$    0.285(2)$\cr
$2.0$&$  0.2577(11)$&$    0.259(3)$&$   0.2581(3)$&$    0.256(3)$&$    0.276(4)$\cr
$3.0$&$  0.2582(11)$&$    0.255(4)$&$   0.2589(4)$&$    0.253(4)$&$    0.261(5)$\cr
$4.0$&$   0.2560(9)$&$    0.262(5)$&$   0.2585(4)$&$    0.261(5)$&    0.262(6)\cr
$5.0$&$  0.2546(11)$&$    0.262(7)$&$   0.2592(4)$&$    0.261(7)$&$    0.265(8)$\cr
\noalign{\smallskip\hrule\smallskip}
\noalign{\bigskip}
$F$ \cr
\noalign{\smallskip\hrule\smallskip}
 $J(\times 10^{-4})$ & $\el$ & $\et$ & $\es$ & $\pt$ & $\ps$ \cr
\noalign{\smallskip\hrule\smallskip\hrule\smallskip}
$0.0$&$    0.290(4)$&$    0.282(3)$&$    0.286(3)$&    ***&    ***\cr
$0.5$&$    0.289(4)$&$    0.286(4)$&$    0.287(4)$&    \#\#\#&$    0.412(5)$\cr
$1.0$&$    0.279(4)$&$    0.291(5)$&$    0.285(3)$&$    0.231(5)$&$    0.312(5)$\cr
$2.0$&$    0.280(5)$&$    0.292(6)$&$    0.287(3)$&$    0.274(6)$&$    0.305(5)$\cr
$3.0$&$    0.272(5)$&$    0.290(6)$&$    0.290(4)$&$    0.284(7)$&$    0.292(6)$\cr
$4.0$&$    0.257(5)$&$    0.292(7)$&$    0.283(3)$&$    0.290(7)$&    0.287(7)\cr
$5.0$&$    0.245(6)$&$    0.298(10)$&$    0.287(4)$&$    0.297(10)$&$    0.293(9)$\cr
\noalign{\smallskip\hrule\smallskip}
}
}$$ 
\centerline{$L=32, \kappa=0.47$} 
\bigskip
\bigskip
\centerline{\cour Table I}
\vfill
\eject

$$\vbox{
\halign{\hfil#\hfil&&\quad\hfil#\cr
$\Sigma$ \cr
\noalign{\smallskip\hrule\smallskip}
 $J(\times 10^{-4})$ & $\el$ & $\et$ & $\es$ & $\pt$ & $\ps$ \cr
\noalign{\smallskip\hrule\smallskip\hrule\smallskip}
$0.0$&$   0.2050(9)$&$   0.2029(8)$&$   0.2040(5)$&    ***&    ***\cr
$0.5$&$   0.2042(8)$&$   0.2052(10)$&$   0.2040(5)$&    \#\#\#&$   0.3545(6)$\cr
$1.0$&$   0.2041(9)$&$  0.2046(13)$&$   0.2036(5)$&$    0.213(3)$&$  0.2276(12)$\cr
$2.0$&$   0.2057(10)$&$  0.2031(19)$&$   0.2047(7)$&$    0.195(2)$&$    0.218(3)$\cr
$3.0$&$  0.2003(13)$&$    0.214(3)$&$   0.2053(5)$&$    0.210(3)$&$    0.225(4)$\cr
$4.0$&$  0.1996(16)$&$    0.214(4)$&$   0.2053(6)$&$    0.212(4)$&    0.220(5)\cr
$5.0$&$  0.1975(17)$&$    0.214(4)$&$   0.2062(7)$&$    0.212(4)$&$    0.222(5)$\cr
\noalign{\smallskip\hrule\smallskip}
\noalign{\bigskip}
$F$ \cr
\noalign{\smallskip\hrule\smallskip}
 $J(\times 10^{-4})$ & $\el$ & $\et$ & $\es$ & $\pt$ & $\ps$ \cr
\noalign{\smallskip\hrule\smallskip\hrule\smallskip}
$0.0$&$    0.229(4)$&$    0.223(4)$&$    0.226(3)$&    ***&    ***\cr
$0.5$&$    0.232(4)$&$    0.232(5)$&$    0.231(4)$&    \#\#\#&$    0.387(7)$\cr
$1.0$&$    0.225(5)$&$    0.229(5)$&$    0.227(4)$&$    0.165(4)$&$    0.250(5)$\cr
$2.0$&$    0.222(5)$&$    0.226(5)$&$    0.225(4)$&$    0.202(5)$&$    0.238(6)$\cr
$3.0$&$    0.197(5)$&$    0.244(5)$&$    0.224(3)$&$    0.236(5)$&$    0.244(5)$\cr
$4.0$&$    0.191(6)$&$    0.237(6)$&$    0.220(4)$&$    0.234(6)$&    0.234(5)\cr
$5.0$&$    0.181(5)$&$    0.241(6)$&$    0.223(4)$&$    0.239(6)$&$    0.239(6)$\cr
\noalign{\smallskip\hrule\smallskip}
}
}$$ 
\vbox{\hfil $L=32, \kappa=0.462$ \hfil} 
\bigskip
\bigskip
\centerline{\cour Table II}
\vfill
\eject

$$\vbox{
\halign{\hfil#\hfil&&\quad\hfil#\cr
$\Sigma$ \cr
\noalign{\smallskip\hrule\smallskip}
 $J(\times 10^{-4})$ & $\el$ & $\et$ & $\es$ & $\pt$ & $\ps$ \cr
\noalign{\smallskip\hrule\smallskip\hrule\smallskip}
$0.0$&$   0.2575(9)$&$   0.2576(9)$&$   0.2576(2)$&    ***&    ***\cr
$0.5$&$  0.2585(13)$&$    0.256(4)$&$   0.2578(5)$&$    0.255(4)$&$    0.278(5)$\cr
$1.0$&$   0.2565(7)$&$    0.264(4)$&$ 0.25774(20)$&$    0.263(4)$&$    0.272(6)$\cr
$3.0$&$   0.2549(7)$&$    0.268(8)$&$   0.2583(3)$&$    0.268(8)$&$    0.270(9)$\cr
\noalign{\smallskip\hrule\smallskip}
\noalign{\bigskip}
$F$ \cr
\noalign{\smallskip\hrule\smallskip}
 $J(\times 10^{-4})$ & $\el$ & $\et$ & $\es$ & $\pt$ & $\ps$ \cr
\noalign{\smallskip\hrule\smallskip\hrule\smallskip}
$0.0$&$    0.288(4)$&$    0.286(5)$&$    0.287(4)$&    ***&    ***\cr
$0.5$&$    0.286(6)$&$    0.290(8)$&$    0.290(6)$&$    0.264(8)$&$    0.312(9)$\cr
$1.0$&$    0.274(4)$&$    0.294(7)$&$    0.284(3)$&$    0.290(7)$&$    0.300(8)$\cr
$3.0$&$    0.247(5)$&$   0.300(10)$&$    0.285(4)$&$   0.300(10)$&$   0.298(10)$\cr
\noalign{\smallskip\hrule\smallskip}
}
}$$ 
\vbox{\hfil $L=48, \kappa=0.47$ \hfil} 
\bigskip
\bigskip
\centerline{\cour Table III}
\vfill
\eject

$$\vbox{
\halign{\hfil#\hfil&&\quad\hfil#\cr
$\Sigma$ \cr
\noalign{\smallskip\hrule\smallskip}
 $J(\times 10^{-4})$ & $\el$ & $\et$ & $\es$ & $\pt$ & $\ps$ \cr
\noalign{\smallskip\hrule\smallskip\hrule\smallskip}
$0.0$&$   0.2037(5)$&$   0.2023(6)$&$   0.2030(3)$&    ***&    ***\cr
$0.5$&$  0.2023(11)$&$  0.2049(18)$&$   0.2028(6)$&$    0.203(2)$&    \#\#\#\cr
$2.0$&$  0.2002(15)$&$    0.204(6)$&$   0.2052(5)$&$    0.203(6)$&$    0.206(6)$\cr
$3.0$&$    0.140(5)$&$    0.846(9)$&$   0.1991(9)$&$    0.204(8)$&$    0.205(9)$\cr
\noalign{\smallskip\hrule\smallskip}
\noalign{\bigskip}
$F$ \cr
\noalign{\smallskip\hrule\smallskip}
 $J(\times 10^{-4})$ & $\el$ & $\et$ & $\es$ & $\pt$ & $\ps$ \cr
\noalign{\smallskip\hrule\smallskip\hrule\smallskip}
$0.0$&$    0.229(3)$&$    0.225(4)$&$    0.227(3)$&    ***&    ***\cr
$0.5$&$    0.220(6)$&$    0.228(7)$&$    0.224(5)$&$    0.197(6)$&    \#\#\#\cr
$2.0$&$    0.188(7)$&$    0.236(8)$&$    0.230(5)$&$    0.235(8)$&$    0.231(7)$\cr
$3.0$&$    0.113(9)$&$     1.13(3)$&$    0.205(6)$&$   0.229(11)$&$   0.225(11)$\cr
\noalign{\smallskip\hrule\smallskip}
}
}$$ 
\vbox{\hfil $L=48, \kappa=0.462$ \hfil} 
\bigskip
\bigskip
\centerline{\cour Table IV}
\vfill
\eject

$$\vbox{
\halign{\hfil#\hfil&&\quad\hfil#\cr
 $\Sigma$ \cr
 \noalign{\smallskip\hrule\smallskip}
 $\kappa$ & $\el$ & $\et$ & $\es$ \cr
\noalign{\smallskip\hrule\smallskip\hrule\smallskip}
$0.460$&$   0.1841(12)$&$   0.1831(12)$&$   0.1836(7)$\cr
$0.462$&$   0.2025(6)$&$    0.2031(6)$&$    0.2028(3)$\cr
$0.470$&$   0.2580(8)$&$    0.2569(10)$&$   0.2574(2)$\cr
\noalign{\smallskip\hrule\smallskip}
\noalign{\bigskip}
 $F$ \cr
 \noalign{\smallskip\hrule\smallskip}
 $\kappa$ & $\el$ & $\et$ & $\es$ \cr
\noalign{\smallskip\hrule\smallskip\hrule\smallskip}
$0.460$&$   0.206(6)$&$    0.205(7)$&$   0.206(7)$\cr
$0.462$&$   0.225(4)$&$    0.223(4)$&$   0.224(4)$\cr
$0.470$&$   0.288(5)$&$    0.284(5)$&$   0.286(4)$\cr
\noalign{\smallskip\hrule\smallskip}
}
}$$
\vbox{\hfil $L=64, J=0.0$ \hfil} 
\bigskip
\bigskip
\centerline{\cour Table V}
\vfill
\eject


$$\vbox{
\halign{\hfil#\hfil&&\quad\hfil#\cr
 $J(\times 10^{-4})$ & $\langle\varphi^0\rangle$ at $\kappa=0.462$ & $\langle\varphi^0\rangle$ at $\kappa=0.47$ \cr
\noalign{\smallskip\hrule\smallskip\hrule\smallskip}
 0.0& $-$0.0009(17)& $-$0.0016(14)\cr
 0.5&  0.0811(15)&  0.1209(17)\cr
 1.0&  0.1486(14)&    0.208(2)\cr
 2.0&  0.2339(13)&  0.3047(15)\cr
 3.0&  0.2662(11)&  0.3405(12)\cr
 4.0&   0.2851(8)&   0.3557(9)\cr
 5.0&   0.2951(9)&   0.3644(9)\cr
\noalign{\smallskip\hrule\smallskip}
}
}$$ 
\vbox{\hfil $L=32,\langle\varphi^0\rangle$ \hfil} 

\bigskip

$$\vbox{
\halign{\hfil#\hfil&&\quad\hfil#\cr
 $J(\times 10^{-4})$ & $\langle\varphi^0\rangle$ at $\kappa=0.462$ & $\langle\varphi^0\rangle$ at $\kappa=0.47$ \cr
\noalign{\smallskip\hrule\smallskip\hrule\smallskip}
 0.0&  0.0002(13)&  0.0018(18)\cr
 0.5&    0.201(2)&    0.279(2)\cr
 1.0&    \%\%\%&  0.3372(13)\cr
 2.0&   0.2940(7)&    \%\%\%\cr
 3.0&   0.3045(8)&   0.3726(7)\cr
\noalign{\smallskip\hrule\smallskip}
}
}$$ 
\vbox{\hfil $L=48,\langle\varphi^0\rangle$ \hfil} 

\bigskip
\bigskip
\centerline{\cour Table VI}
\vfill
\eject

$$\vbox{
\halign{\hfil#\hfil&&\quad\hfil#\cr
 $J(\times 10^{-4})$ & $\chi_{||}$ at $\kappa=0.462$ & $\chi_{||}$ at $\kappa=0.47$ \cr
\noalign{\smallskip\hrule\smallskip\hrule\smallskip}
 0.0&    1710.10(5)&    2545.59(7)\cr
 0.5&   1534.67(12)&   2166.94(17)\cr
 1.0&   1183.82(15)&     1508.6(3)\cr
 2.0&   523.61(18)&     522.3(3)\cr
 3.0&   250.24(17)&     213.9(2)\cr
 4.0&   136.08(12)&   106.70(18)\cr
 5.0&   97.89(15)&   77.62(17)\cr
\noalign{\smallskip\hrule\smallskip}
}
}$$ 
\vbox{\hfil $L=32,\chi_{||}$ \hfil} 

\bigskip

$$\vbox{
\halign{\hfil#\hfil&&\quad\hfil#\cr
 $J(\times 10^{-4})$ & $\chi_{||}$ at $\kappa=0.462$ & $\chi_{||}$ at $\kappa=0.47$ \cr
\noalign{\smallskip\hrule\smallskip\hrule\smallskip}
 0.0&    5438.72(6)&   8258.40(15)\cr
 0.5&     2311.5(5)&     2447.5(7)\cr
 1.0&    \%\%\%&     589.7(4)\cr
 2.0&     148.9(2)&    \%\%\%\cr
 3.0&     71.5(2)&     72.9(3)\cr
\noalign{\smallskip\hrule\smallskip}
}
}$$ 
\vbox{\hfil $L=48,\chi_{||}$ \hfil} 

\bigskip
\bigskip
\centerline{\cour Table VII}
\vfil
\eject

\epsfysize=0.8\vsize
\centerline{\epsfbox{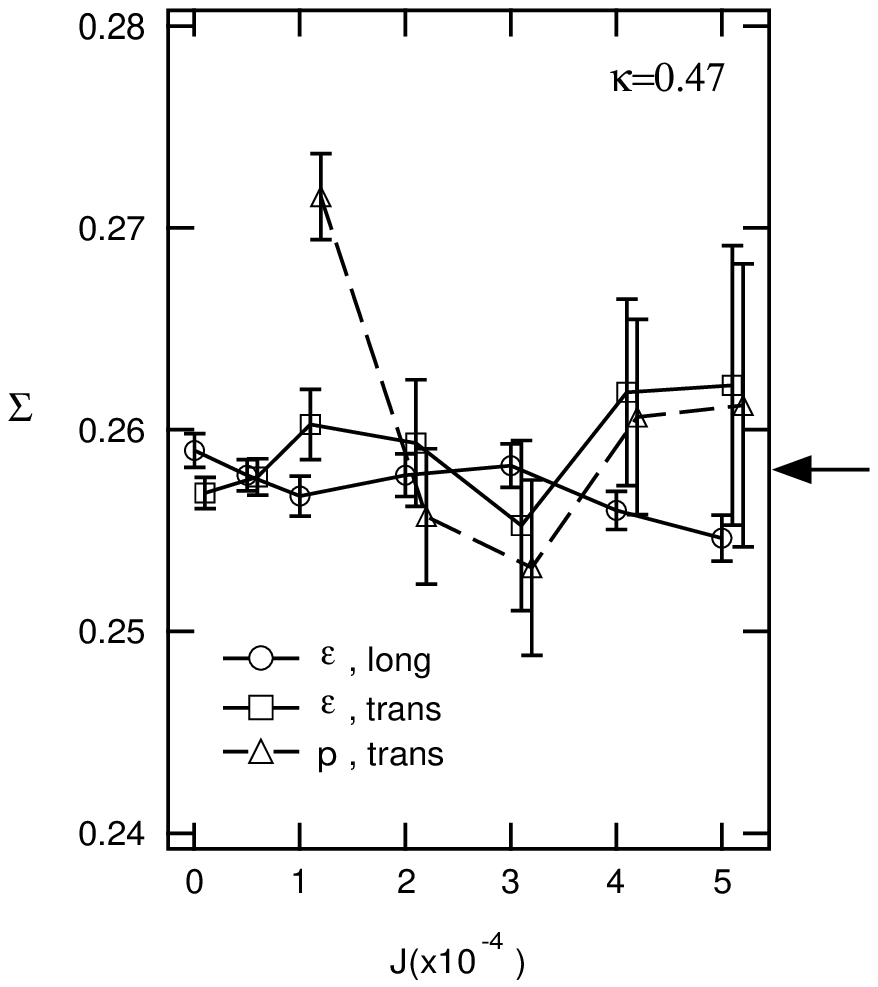}}
\vskip2em
\centerline{{\cour Fig.1}}
\vfil\eject

\epsfysize=0.8\vsize
\centerline{\epsfbox{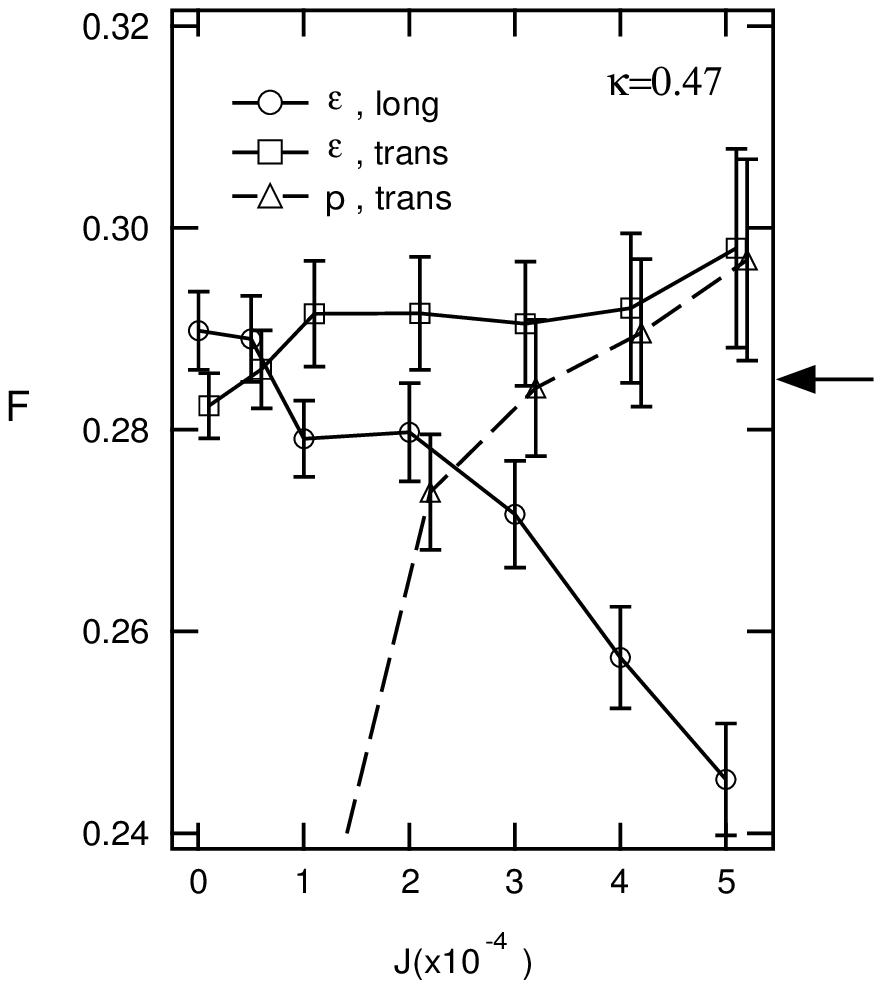}}
\vskip2em
\centerline{{\cour Fig.2}}
\vfil\eject

\epsfysize=0.8\vsize
\centerline{\epsfbox{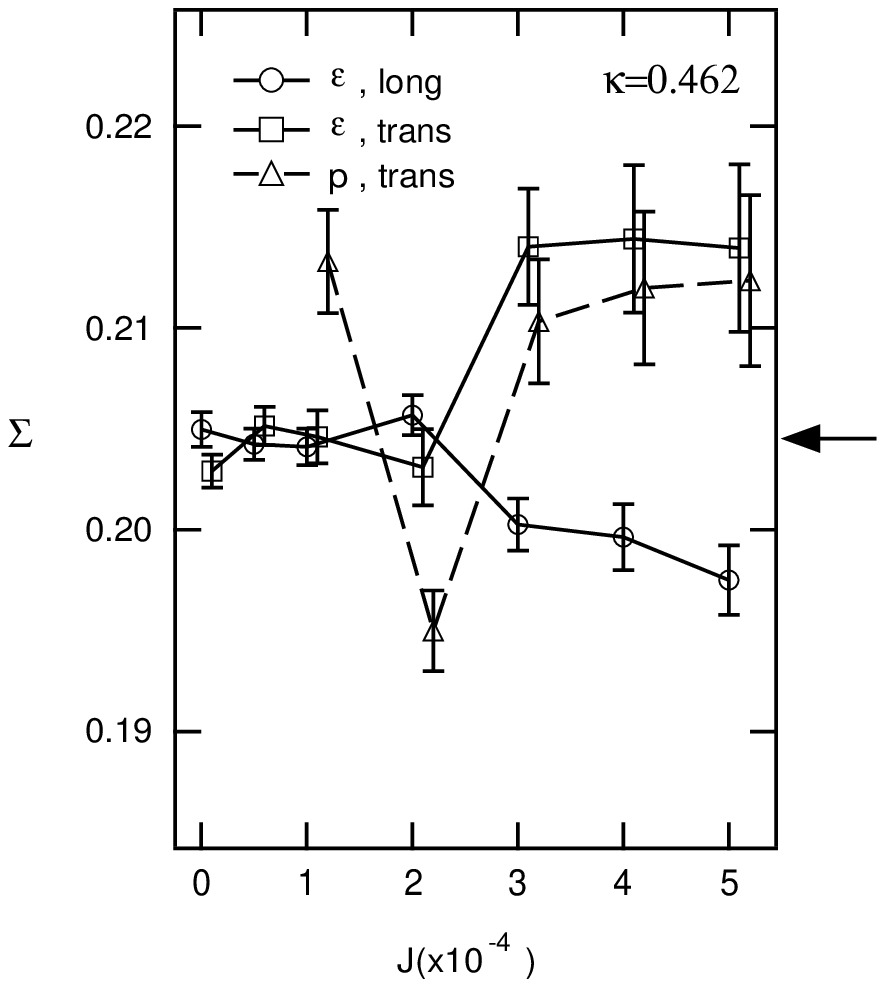}}
\vskip2em
\centerline{{\cour Fig.3}}
\vfil\eject

\epsfysize=0.8\vsize
\centerline{\epsfbox{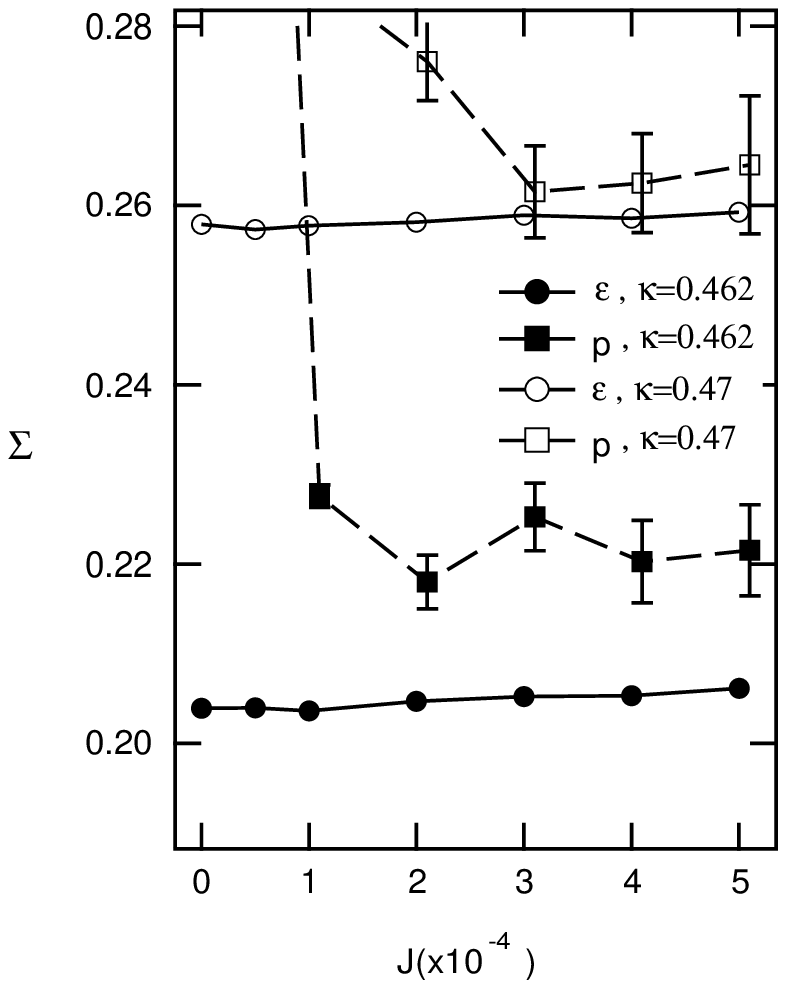}}
\vskip2em
\centerline{{\cour Fig.4}}
\vfil\eject

\epsfysize=0.8\vsize
\centerline{\epsfbox{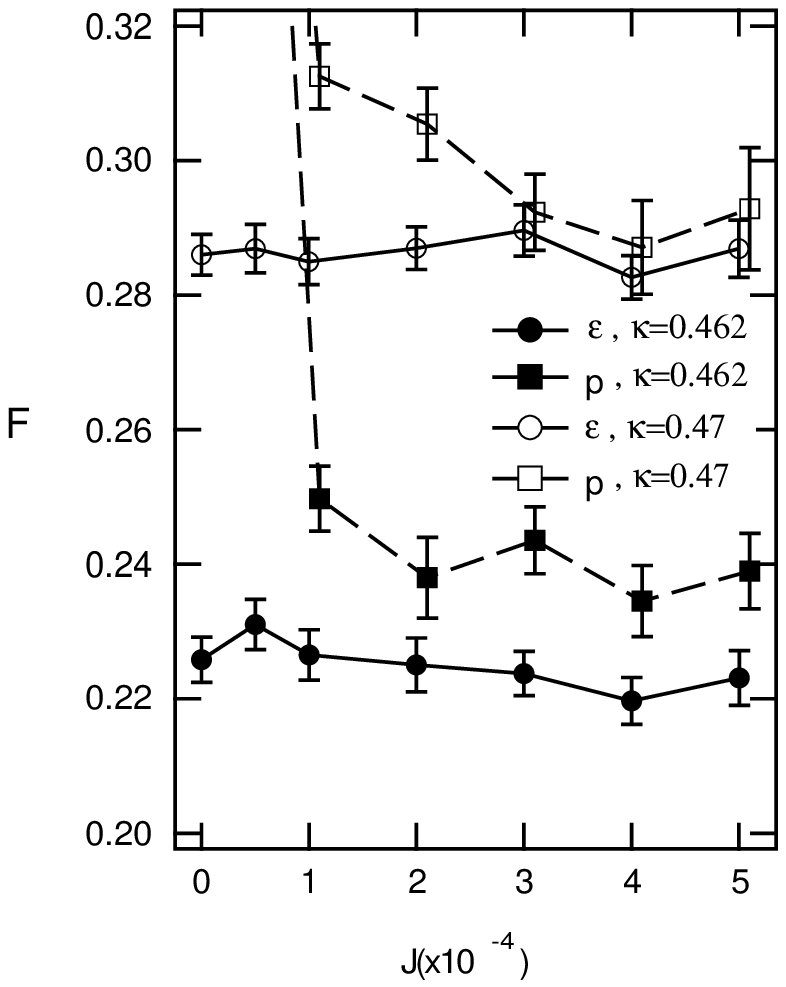}}
\vskip2em
\centerline{{\cour Fig.5}}
\vfil\eject

\epsfysize=0.8\vsize
\centerline{\epsfbox{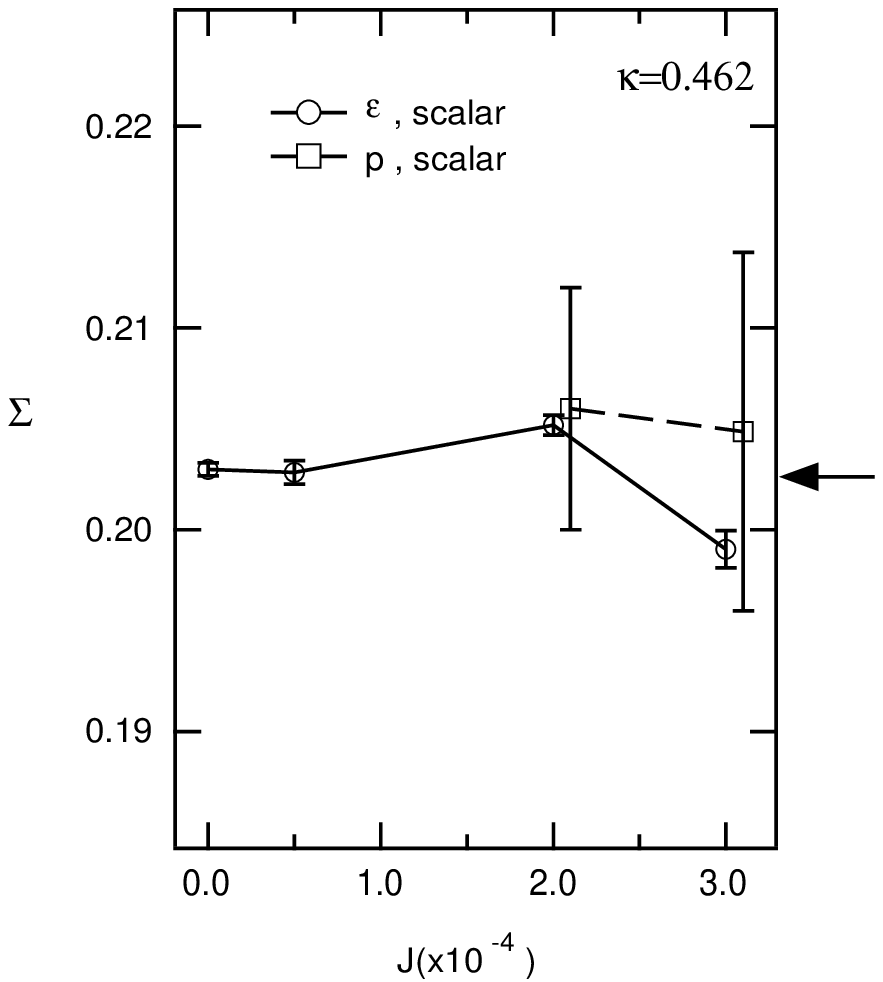}}
\vskip2em
\centerline{{\cour Fig.6}}
\vfil\eject

\epsfysize=0.8\vsize
\centerline{\epsfbox{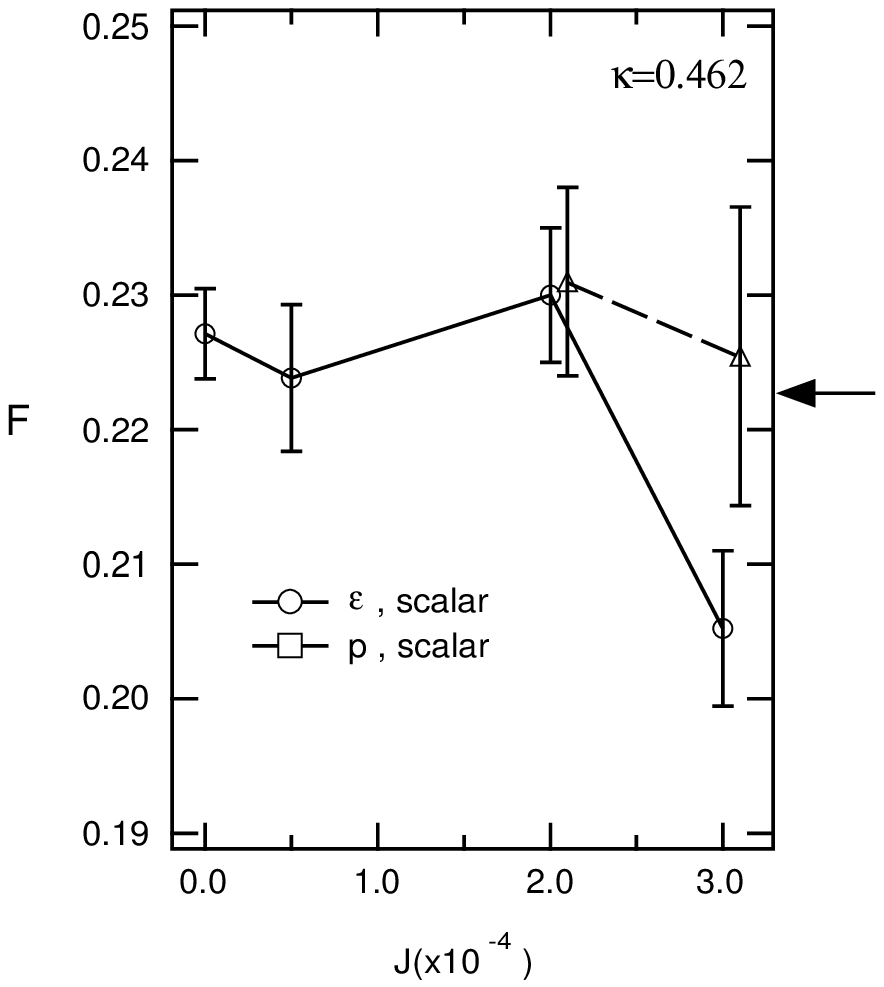}}
\vskip2em
\centerline{{\cour Fig.7}}
\vfil\eject

\epsfysize=0.8\vsize
\centerline{\epsfbox{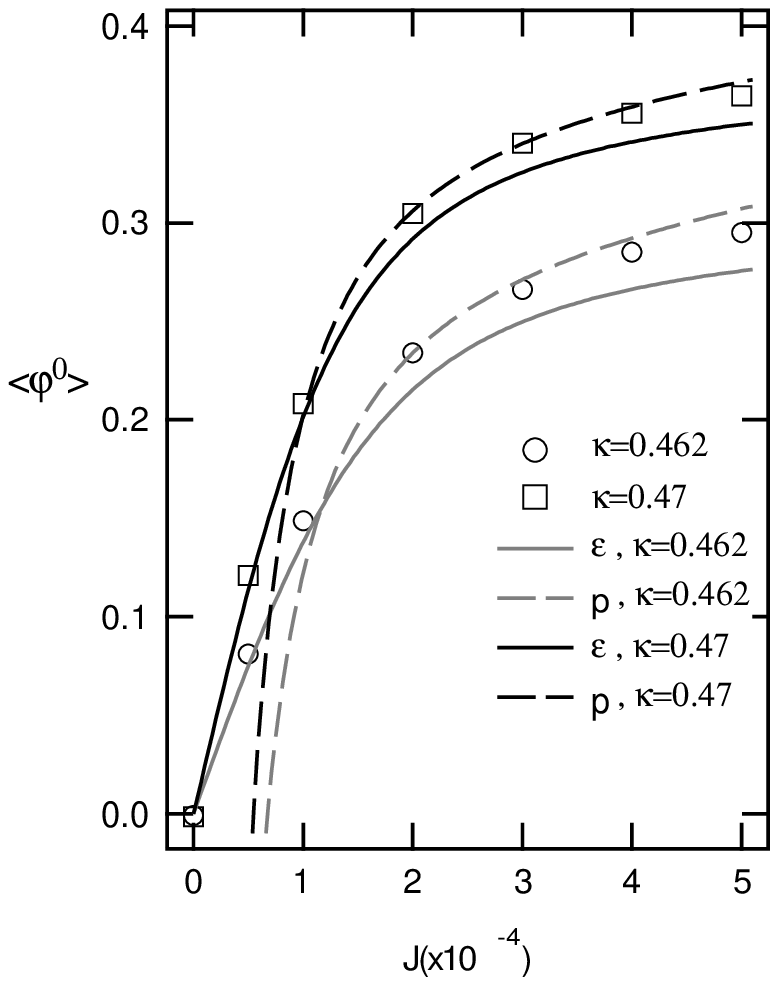}}
\vskip2em
\centerline{{\cour Fig.8}}
\vfil\eject

\epsfysize=0.8\vsize
\centerline{\epsfbox{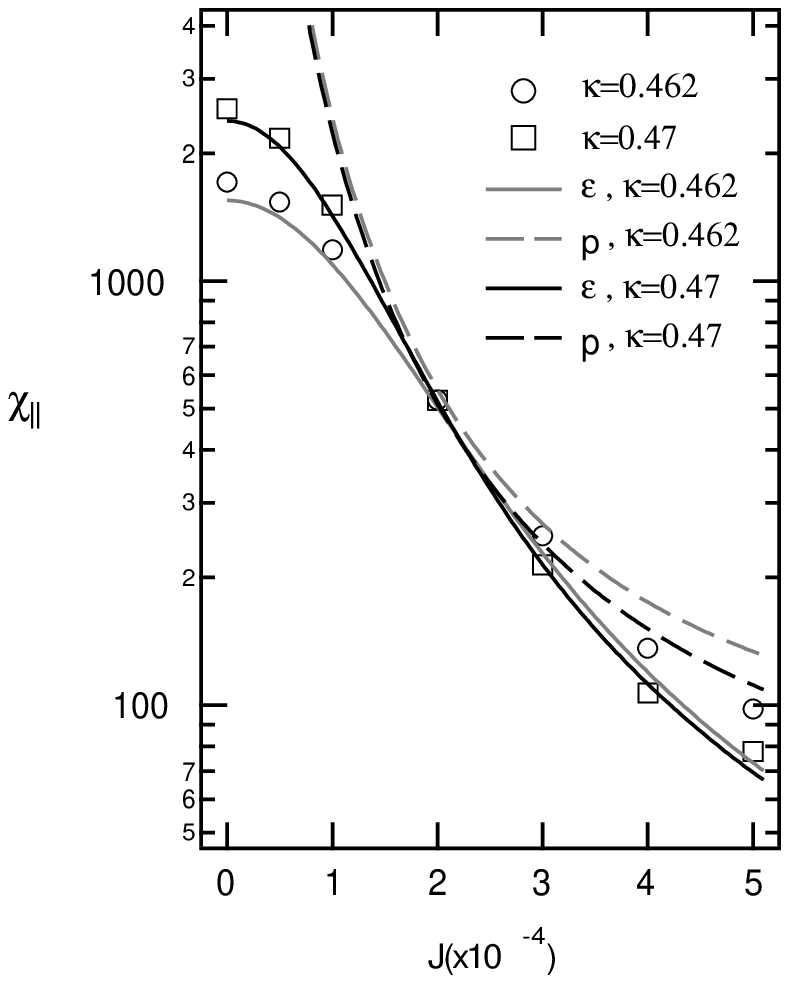}}
\vskip2em
\centerline{{\cour Fig.9}}
\vfil\eject

\bye